\documentclass[showkeys,twocolumn]{revtex4}
\usepackage{epsfig}
\usepackage{graphicx}
\usepackage{amsmath}
\begin{document}
\title {The Evolution of Complexity in Social Organization - A Model Using Dominance-Subordinate Behaviour in Two Social Wasp Species} 
\author{Anjan K. Nandi$^{1,}$\footnote{Corresponding author:\\Email: anjanandi@gmail.com}, Anindita Bhadra $^2$, Annagiri Sumana$^2$ Sujata A. Deshpande$^3$ and Raghavendra Gadagkar$^{1,4}$}
\affiliation{
$^{1}$ Centre for Ecological Sciences, Indian Institute of Science, Bangalore 560012, India\\
$^{2}$ Department of Biological Sciences, Indian Institute of Science Education and Research - Kolkata, Mohanpur 741252, India\\
$^{3}$ St. Xavier's College (Autonomous), 5 Mahapalika Marg, Mumbai 400001, India\\
$^4$ Evolutionary and Organismal Biology Unit, Jawaharlal Neheru Centre for Advanced Scientific Research, Jakkur, Bangalore 560004, India}
\begin{abstract}
Dominance and subordinate behaviours are important ingredients in the social organizations of group living animals. Behavioural observations on the two eusocial species \textit{Ropalidia marginata} and \textit{Ropalidia cyathiformis} suggest varying complexities in their social systems. The queen of \textit{R. cyathiformis} is an aggressive individual who usually holds the top position in the dominance hierarchy although she does not necessarily show the maximum number of acts of dominance, while the \textit{R. marginata} queen rarely shows aggression and usually does not hold the top position in the dominance hierarchy of her colony. In \textit{R. marginata}, more workers  are involved in dominance-subordinate interactions as compared to \textit{R. cyathiformis}. These differences are reflected in the distribution of dominance-subordinate interactions among the hierarchically ranked individuals in both the species. The percentage of dominance interactions decrease gradually with hierarchical ranks in \textit{R. marginata} while in \textit{R. cyathiformis} it first increases and then decreases. We use an agent-based model to investigate the underlying mechanism that could give rise to the observed patterns for both the species. The model assumes, besides some non-interacting individuals, that the interaction probabilities of the agents depend on their pre-differentiated winning abilities. Our simulations show that if the queen takes up a strategy of being involved in a moderate number of dominance interactions, one could get the pattern similar to \textit{R. cyathiformis}, while taking up the strategy of very low interactions by the queen could lead to the pattern of \textit{R. marginata}. We infer that both the species follow a common interaction pattern, while the differences in their social organization are due to the slight changes in queen as well as worker strategies. These changes in strategies are expected to accompany the evolution of more complex societies from simpler ones.
\end{abstract}
\keywords{Ropalidia, eusociality, hierarchy, winning ability, queen strategy}
\maketitle

\section{Introduction} 
Group living animals display fascinating diversity in their social systems. Even within the class of the `truly' social or eusocial species, there exists a considerable degree of variation in their social development. Eusociality, which is characterized by reproductive division of labour, cooperative brood care, and overlap of generations, is mainly seen in ants, bees, wasps and  termites \cite{Wilson.71}. The presence or absence of morphologically distinguishable reproductive and non-reproductive castes is used to further subdivide eusocial species into highly and primitively eusocial respectively \cite{Wilson.71}. Honeybees and many species of ants are examples of advanced or highly eusocial societies; their colonies are usually large, consisting of thousands of workers and one or a small subset of individuals who are queens. In typical highly eusocial species like the honeybees the queens use pheromones to maintain reproductive monopoly over their workers. The workers show well-defined division of labour, their activities are self-organized or being regulated by the non-interactive queen through chemicals, rather than physical interactions. By contrast, the primitively eusocial species maintain comparatively smaller colonies with fewer workers, and the queens are generally highly interactive \cite{Wilson.71}. The primitively eusocial queen was previously believed to use aggression to suppress reproduction in the workers, and maintain worker activities in the colony \cite{Reev.Gamb.83, Reev.Gamb.87, It.Hig.91, It.93, Mon.Peet.99} though the notion of work regulation using aggression has been challenged \cite{Premnath.etal.95, Jha.etal.06}. The presence of division of labour \cite{Ng.Gad.98} and queen pheromones \cite{Sled.etal.01, Dapp.etal.07, Sum.etal.08, Bh.etal.10} have been reported in some primitively eusocial species also, but it is still generally believed that, to successfully control the large number of workers in highly eusocial colonies, the non-interactive queens exhibit more complex control systems than their highly interactive primitively eusocial counterparts.
 
Interactions between individuals are critical in social organization, and dominance-subordinate interactions contribute largely to the total interactions observed in insect societies \cite{Wilson.71}. Social dominance hierarchies based on such agonistic interactions is a usual way of ranking individuals. Such dominance hierarchies have long been known to exist in group-living animals, for example in birds \cite{TSEb.35}, cattle \cite{Sch.Foh.55}, fish \cite{Low.56}, primates \cite{Bal.71, Sm.99, Sap.93} and in other beasts \cite{Tyl.72, Clutt.82}. As early as in the 1930’s, some correlates of dominance were identified for the vertebrates and this made it possible to modify the existing social orders by experimental manipulations \cite{A.etal.39}. Pardi showed that the very idea of dominance could be extended to the invertebrates also, and it is his pioneering work in the next decade that revealed the existence of a similar kind of social hierarchy in the primitively eusocial wasp \textit{Polistes dominulus}, then known as \textit{Polistes gallicus} (L.) \cite{Pardi.48}. 

\textit{Ropalidia marginata} is characterized as a primitively eusocial wasp species due to the absence of a morphologically distinguishable queen caste, but unlike in other such species, the queen is usually a meek and docile individual who rarely participates in any dominance interactions with any of her nestmates \cite{Chandra.Gadagkar.91, Gadagkar.01}. As a result, she does not usually hold the top position in the dominance hierarchy of her colony \cite{Sum.Gad.01}. On the other hand, \textit{Ropalidia cyathiformis} is another primitively eusocial species closely related to \textit{R. marginata}, where the queen is aggressive, frequently indulges in dominance interactions with others and usually occupies the top position in the dominance hierarchy \cite{Kar.Gad.02, Kar.Gad.03}. The presence of a de-centralized work regulation mechanism, age polyethism, a non-aggressive queen who uses pheromones to regulate worker reproduction and a pre-determined succession hierarchy makes the social organization of \textit{R. marginata} more complex than most other primitively eusocial societies, including \textit{R. cyathiformis} \cite{Premnath.etal.95, Ng.Gad.98, Sum.etal.08, Bh.etal.10, Bruyndonckx.etal.06, Lamba.etal.07, Bh.etal.07, Bh.Gad.08, Bang.Gad.12}. These two species together present an interesting scenario where it is possible to study differences in social organization in closely related species that might provide hints towards the evolution of complexity in social systems. 

In this paper, we study the distribution of dominance and subordinate interactions among the hierarchically ranked individuals in \textit{R. marginata} and \textit{R. cyathiformis}. We focus on the differences in the dominance and subordinate patterns of the two species that might lead to their different levels of social complexity and thus hope to explore the underlying mechanisms that could delineate these patterns. How does the mechanism differ in these two species? Can we explain the differences between the two species in terms of simple changes in strategies of individual wasps? We use agent-based modeling to check if a common model could explain the behavioural patterns present in these species. Such an exercise could help us to trace the pathway for the evolution of more complex societies from simpler ones. Though the present study was stimulated by observations on these wasp species, the model we introduce and develop is not restricted only to the social insects; with further modifications and additional relevant parameters, we expect our model to be well applicable for even more complex societies including those of vertebrates.
\begin{figure*}
\begin{center}
\includegraphics[width=18cm]{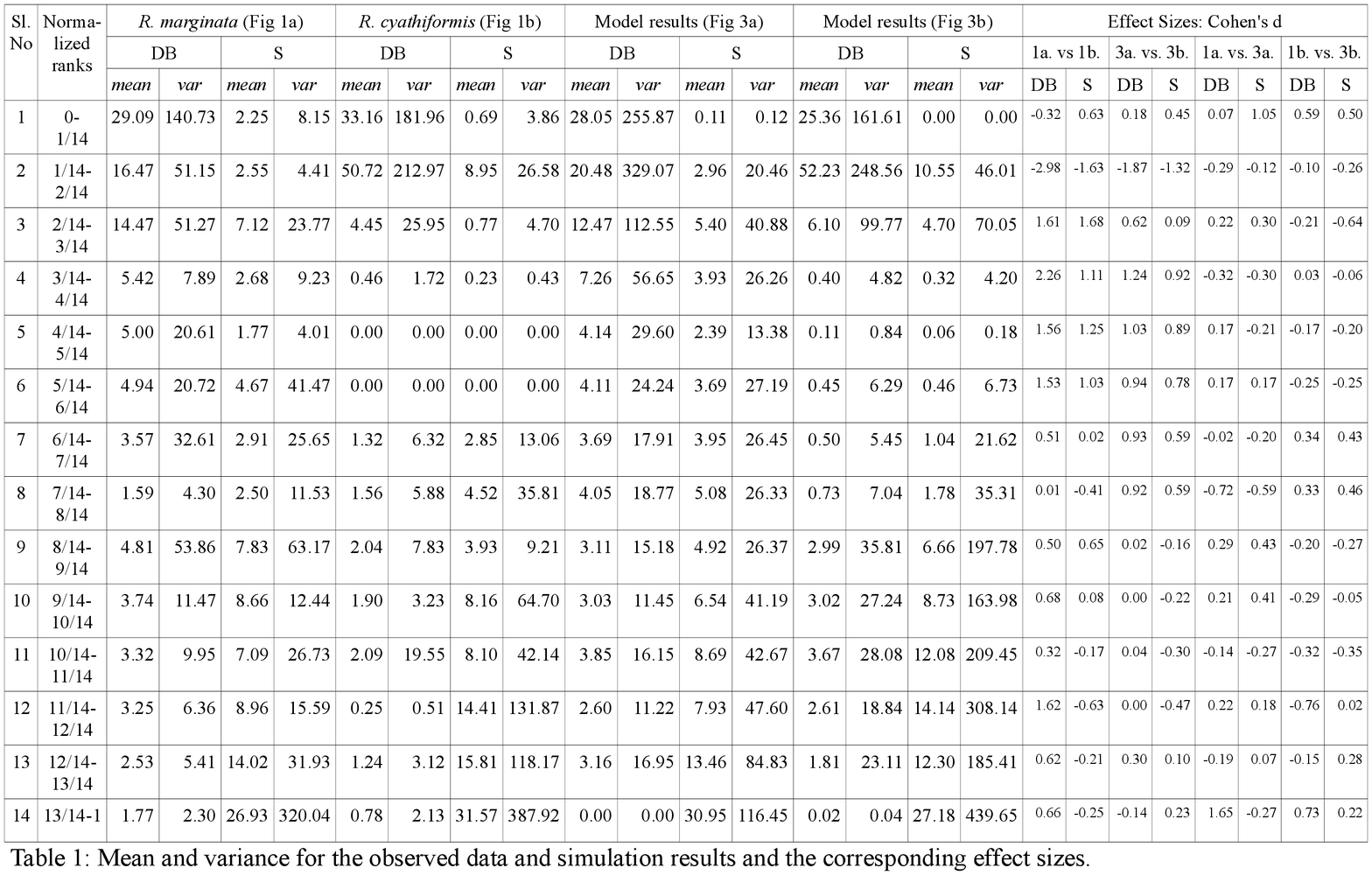}
\end{center}
\end{figure*}

\section{The Experimental Data}
\subsection{Data Collection}
We used data from experiments conducted on nine colonies of \textit{R. marginata} and \textit{R. cyathiformis} each to investigate the patterns of dominance and subordinate behaviours in these species. The colony sizes ranged from $14$ to $59$ adults in \textit{R. marginata}, and from $15$ to $24$ adults in \textit{R. cyathiformis}. The adults on the nest were uniquely marked with spots of Testors\textsuperscript{\textregistered} quick drying enamel paints prior to the observations. Behavioural observations consisted of randomly intermingled `instantaneous scans' (in which a snapshopt of the behavioral state of each individual was recorded) and `all occurrences sessions' (in which every occurrence of a small set of chosen behaviors by any individual was recorded), each session lasting 5 min and followed by a break of 1 min between every session \cite{Gadagkar.01}. Such observations were made for 5 hours each day in two separate blocks of 2h 30min each, over 6 consecutive days in \textit{R. marginata}, yielding 30 hours of data with 96 scans and 204 all occurrences sessions per nest. In case of \textit{R. cyathiformis}, observations were made for $9$ hours in a single day in three separate blocks of $3$ hours each, consisting of $45$ scans and $45$ all occurrences sessions per nest. The queens were identified by observing the egg-laying behaviour.
\begin{figure}
\begin{center}
\includegraphics[width=8cm]{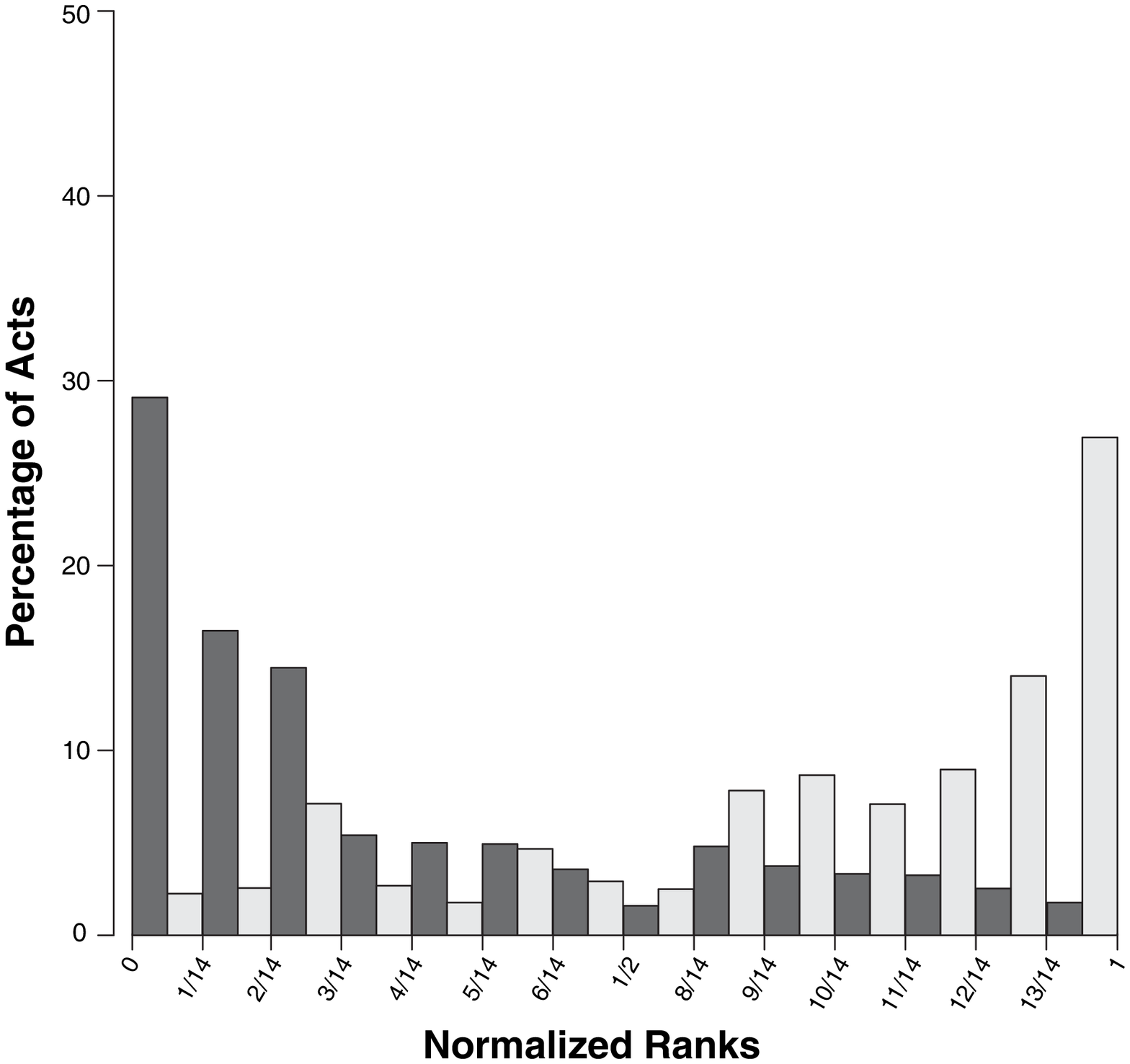}
\includegraphics[width=8cm]{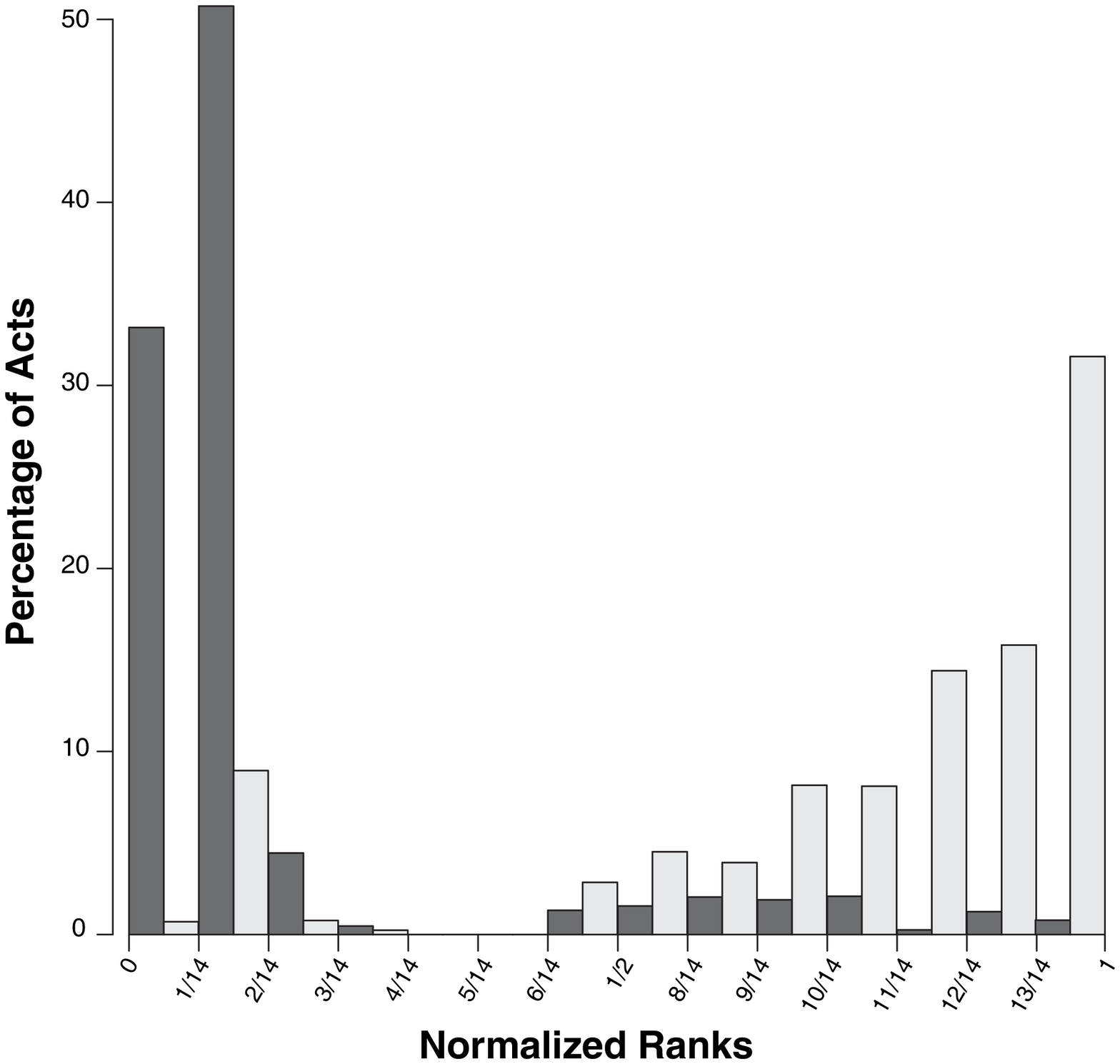}
\end{center}
\caption{Percentage of dominance acts (black bars) and subordinate acts (gray bars) plotted against the normalized ranks for a) 9 colonies of \textit{R. marginata} and b) 9 colonies of \textit{R. cyathiformis}.}
\end{figure}

\subsection{The Dominance Patterns}
All instances of nine different behaviours such as: attack, chase, nibble, peck, crash, sit over another individual, being offered food, aggressive biting and hold another individual by mouth \cite{Chandra.Gadagkar.91, Gadagkar.01} were pooled to calculate the dominance behaviour shown by an individual. The recipient of each of these behaviours was given a score of 1 for computing rates of subordinate behaviour. A dominance hierarchy for each nest was constructed using FDI (frequency based dominance index), which has been shown to be a good index for constructing dominance hierarchies in such wasp species \cite{Bang.etal.10}. In a colony of $n$ individuals, each individual is given an index of dominance $D$ using the following formula:
\begin{equation*}
D=\frac{\sum_{i=1}^nB_i+\sum_{j=1}^m\sum_{i=1}^nb_{ji}+1}{\sum_{i=1}^nL_i+\sum_{j=1}^p\sum_{i=1}^nl_{ji}+1}
\end{equation*}
where $\Sigma_i{B_i}$ denotes the rates at which the focal individual shows dominance behaviour toward her colony members, $\Sigma_{ji}b_{ji}$ denotes the sum of the rates at which all individuals dominated by her show dominance behaviour toward other colony members; $1$ to $m$ are thus individuals who have received aggression from the focal individual. Similarly, $\Sigma_i L_i$ denotes the rates at which the focal individual shows subordinate behaviour toward her colony members, $\Sigma_{ji}l_{ji}$ denotes the sum of the rates at which all individuals who show aggression to the focal individual show subordinate behaviour toward other colony members. Thus $1$ to $p$ are the individuals toward whom the focal individual shows subordinate behaviour. Thus each individual including those who have not shown any dominance-subordinate interactions gets an index of dominance $D$ and the individual with the highest $D$ gets the top position in the dominance hierarchy \cite{Prem.etal.90}. Since this index takes into consideration both the indirect dominance and the indirect subordination shown by the individuals (by means of $b_{ji}$ and $l_{ji}$), showing the most number of dominance behaviours does not guarantee that an individual would hold the topmost position in the hierarchy. We arrange all the individuals of the colony in decreasing order of their value of this FDI index and assign them ranks from one to $n$. Since the colonies have variable number of individuals, in order to pool the data we need to convert the ranks into normalized ranks. This is done by dividing each individual's rank by the total number of individuals in that colony. By doing so, the ranks of all the individuals are scaled between $0$ and $1$. We calculate the percentage of dominance-subordinate behaviour shown by each individual for each colony and plot this against the individual’s normalized rank. For this purpose we divide the scale of normalized ranks ($0$-$1$) into $14$ bins of equal size. For instance, if the $10$th ranked individual in a colony of $30$ individuals shows $5$ acts of dominance behaviour where a total of $20$ acts are recorded in the colony, then $(5/20)*100\%=25\%$ would be added to the $5$th bin, since the normalized rank $10/30=0.33$ lies between $4/14$ and $5/15$. Since the smallest colony had $14$ individuals, in order to ensure that each bin has at least one individual for each nest, we used $14$ bins. Thus in colonies of \textit{R. marginata}, individuals who have normalized hierarchical ranks between $0$ and $1/14$ show $29.09\%$ of the total dominance and $2.25\%$ of the total subordination of the colony, those having ranks between $1/14$ and $2/14$ show around $16.47\%$ of the total dominance and $2.55\%$ of the total subordination and so on (Table 1). We plotted this distribution in form of histogram in figure 1a, where black bars and gray bars represent the dominance and subordinate behaviours respectively. Similar analysis for the nine colonies of \textit{R. cyathiformis} yielded the pattern observed in figure 1b; mean and variance for the bars are shown in Table 1.
\begin{figure*}
\begin{center}
\includegraphics[width=15cm]{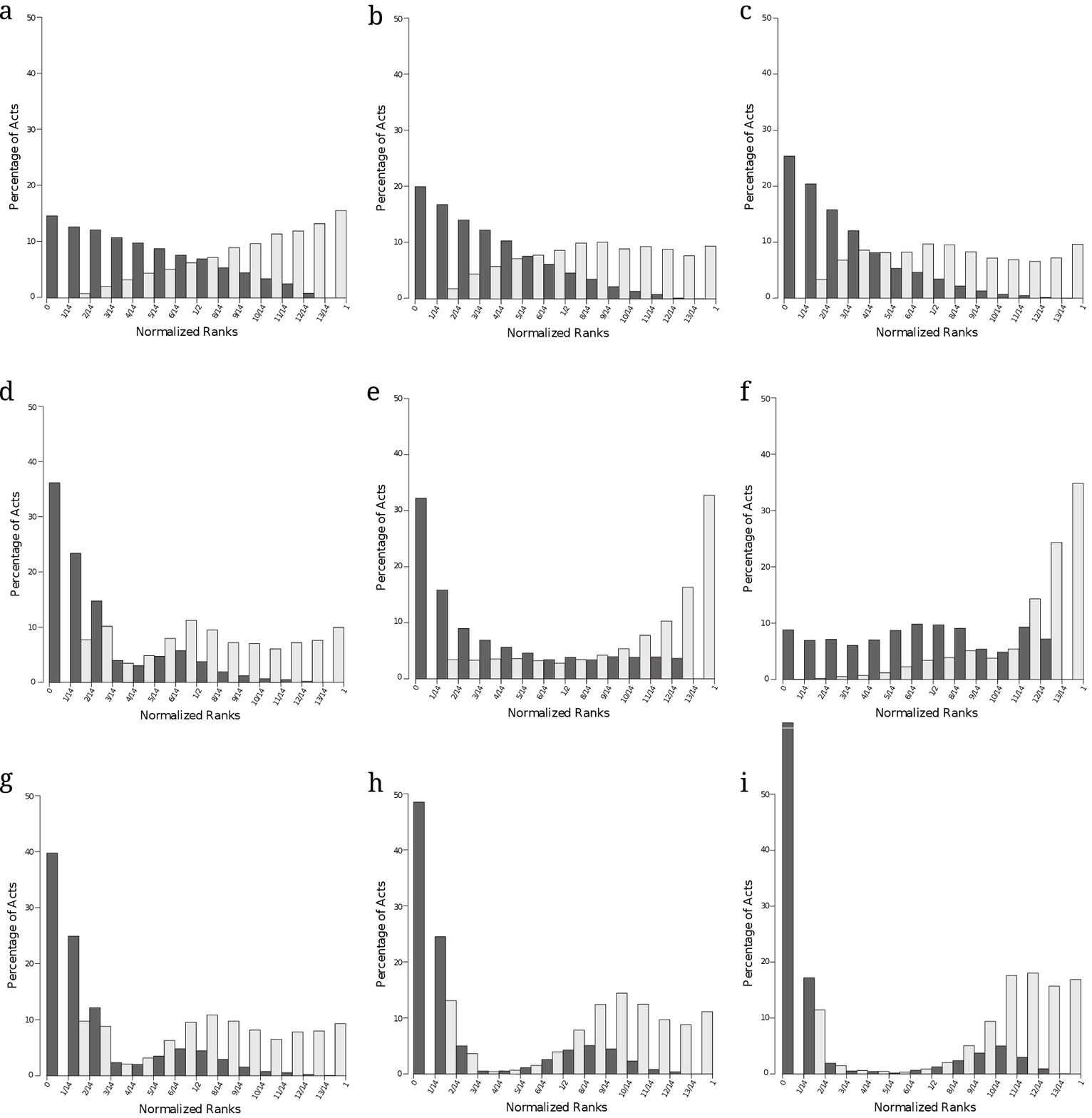}
\end{center}
\caption{Percentage of dominance acts (black bars) and subordinate acts (gray bars) plotted against the normalized ranks obtained by averaging over 100 configurations of 100 interactions with parameters a) $w=0$, $\alpha=0$, $\beta=0$, b) $w=0$, $\alpha=0$, $\beta=0.5$, c) $w=0$, $\alpha=0$, $\beta=1$, d) $w=0$, $\alpha=0$, $\beta=2$, e) $w=0$, $\alpha=0.5$, $\beta=2$, f) $w=0$, $\alpha=1$, $\beta=2$, g)  $w=10$, $\alpha=0$, $\beta=2$, h) $w=30$, $\alpha=0$, $\beta=2$, i) $w=50$, $\alpha=0$, $\beta=2$.}
\end{figure*}

\subsection{The Comparison}
There are striking differences between the dominance-subordinate behaviours shown by the two species. While in all the nine colonies, the \textit{R. cyathiformis} queen always held the top position in the dominance hierarchy, the \textit{R. marginata} queen never held the top position in any of the nine colonies analyzed, her position being different in different colonies, ranging from $4$ to $22$. We compare the behavioural patterns shown by the two species (figure 1a and figure 1b) by means of Kolomogorov-Smirnov two sample test \cite{Sok.Rof.94}. We perform the test for dominance as well as subordinate behaviours separately and found a significant difference at $95\%$ confidence level for both the behaviours. We therefore conclude that the two species show dominance and subordinate patterns different from each other. The same conclusion could be drawn by using Cliff’s delta \cite{Cliff.93, Cliff.96}, a measure of effect size, which represents the degree of overlap between two distributions. We calculated Cliff’s delta of $0.58$ for dominance and $0.10$ for subordinate behaviours. We also compare each bar of one distribution with its corresponding bar in the other distribution by measuring Cohen’s $d$ index \cite{Cohen.88}, another measure of effect size and found high effect sizes for many of the bars; all the details are presented in Table 1. Dominance in \textit{R. marginata} is found to decrease almost consistently with the ranks for the higher ranks (black columns in figure 1a). Tukey multiple comparison test for proportion \cite{Zar.09} confirms significant differences at $95\%$ confidence level between column $1$ and column $2$, and also between column $3$ and column $4$, although the difference between column $2$ and column $3$ was insignificant. For \textit{R. cyathiformis} however, it is very evident from the size of the black columns of figure 1b, that dominance behavior first increases and then decreases. Tukey test also shows significant difference at $95\%$ confidence level between column $1$ and $2$, also between column $2$ and $3$. Since in all the nine colonies of \textit{R. cyathiformis} the queen holds the top position in the hierarchy, this analysis suggests that she may not show maximum amount of dominance but there could be other individuals present in the colony who show more dominance than the queen. From the data we also found the acts of aggression shown by the individual holding the second position in the hierarchy to be numerically more than that of the queen in six out of nine colonies. Statistically, in all the nine colonies, the dominance shown by the second ranking individual is comparable with the queen (Tukey test, no significant difference at $95\%$ confidence level). In both the species, the subordinate behaviours are not distributed equally among the workers, but gradually increase with their ranks for lower ranking individuals (gray bars). The two empty bins in the pattern of \textit{R. cyathiformis} suggest the presence of more non-interacting workers in this species as compared to \textit{R. marginata}. 
\begin{figure}
\begin{center}
\includegraphics[width=8cm]{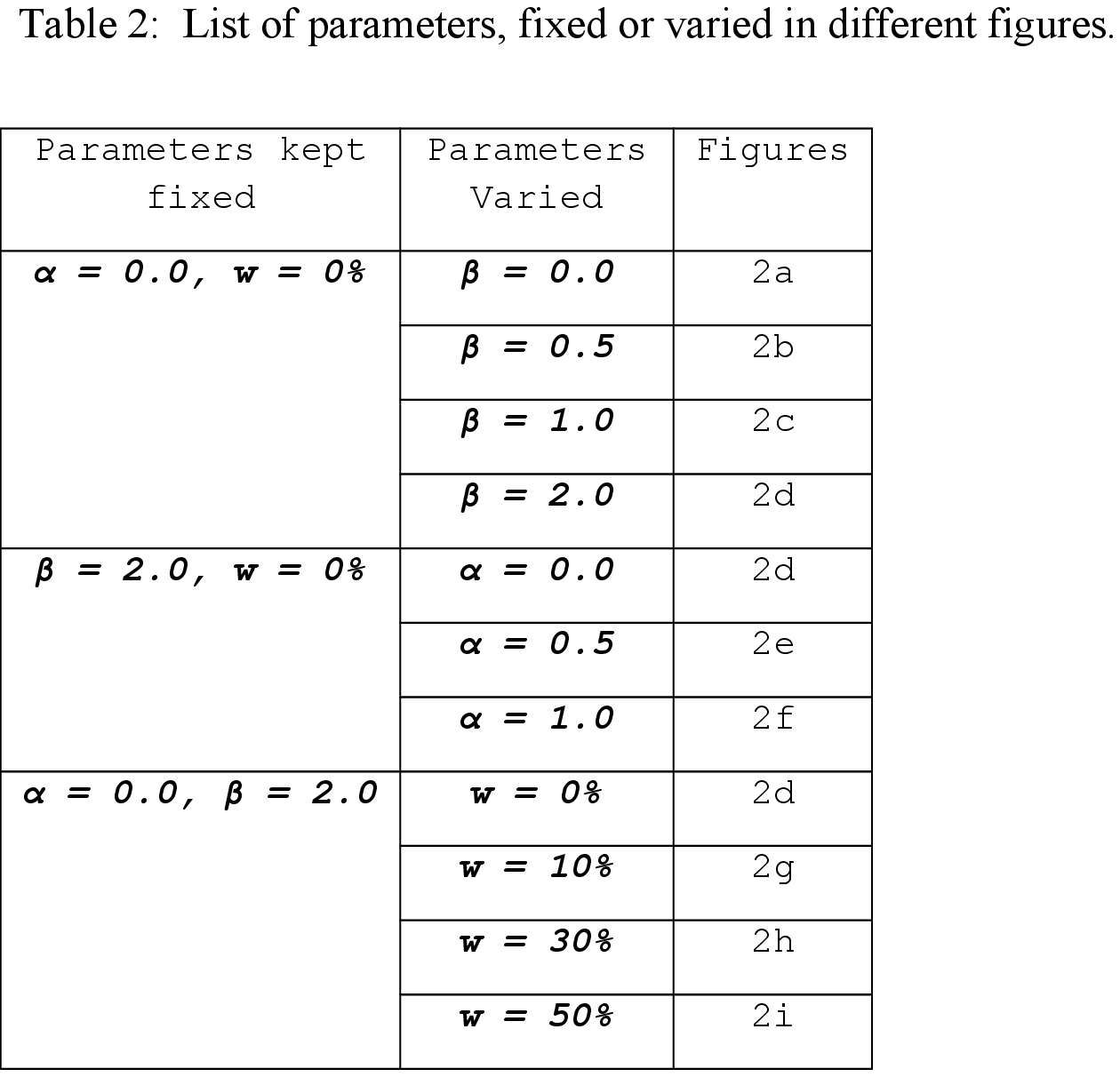}
\end{center}
\end{figure}

\section{The model}
Can simple changes in strategies of the individuals explain the existing differences between the two species? We attempt to build a single model to explain the dynamics of the colonies that could give rise to these patterns. Existing models of dominance patterns can generally be classified into two categories. Self-organized models rely on a reinforcement mechanism that, depending on an individual's previous experiences, increases or decreases its ability to dominate others in an agonistic interaction \cite{Hog.Hos.83, Jag.Seg.92, Ther.etal.95, Bon.etal.96}. On the other hand, Correlational models assume pre-differentiated winning abilities in the individuals and further assume that their hierarchical ranks directly reflect their winning abilities \cite{Chase.86, Bon.etal.99}. Both the models are found to be equally capable of reproducing the dominance-subordinate patterns seen in \textit{Polistes dominulus} \cite{Bon.etal.99}. Our model is closely related to the aforesaid correlational models, though the hierarchical ranks of the individuals do not always strictly follow the ordering of their winning abilities. 

In our model each individual $i$ is characterized by a strength function $x_i$ which determines their winning abilities in an dominance interaction, i.e., if two individuals $i$ and $j$ meet, $i$ will win over $j$ if $x_i>x_j$, $j$ will win over $i$ if $x_j>x_i$, and if $x_i=x_j$, then both the individuals will have equal chances to win over the other. Let the individuals interact with an interaction probability $p_i$. For a certain proportion of individuals, let’s say for $w$, we set $p_i=0$, i.e., $w$ proportion of individuals are non-interacting, and for the rest of the individuals, $p_i$ is a function of their respective strengths $x_i$. The functional relationship between them is expressed by 
\begin{equation*}
p_i=f(x_i)\sim |x_i-\alpha|^{\beta}
\end{equation*}
where $\alpha$ and $\beta$ are parameters for monotonicity and homogeneity respectively (explained in the next section). We take $N=14$ individuals, assign their strength $x_i$ from a uniform random distribution ranging between $0$ and $1$, then determine their interaction probabilities $p_i$ according to the functional relationship described above and with a specific value of $w$, subjected to the normalization condition $\Sigma_i p_i=1$. For each interaction, we choose two individuals at a time according to their $p_i$'s and determine dominant and subordinate according to their $x_i$'s. We allow $100$ such interactions. The dominance hierarchy is then constructed and percentage of interactions shown by each individual is also calculated. The whole process is repeated for $100$ configurations. Then we bin the interaction data together for their respective normalized ranks as we did for the real data. We declare the individual with the highest $x_i$ as the queen for each of the configurations and also track her position in the respective hierarchies.

\section{Results}

\subsection{The effects of parameter}
The effect of the parameters on the model can be seen by the following. We first examine the response of the model for varying values of the homogeneity parameter $\beta$ with $\alpha=0$ and $w=0$. When $\beta$  is $0$, all $p_i$ s become equal, all individuals have equal probability to interact with all others. So the dominance-subordinate pattern reflects only their winning abilities (figure 2a). The inhomogeneity within the interaction probabilities increases as $\beta$ differs from zero. The scenarios are depicted in figure 2b with $\beta=0.5$, in figure 2c with $\beta=1.0$ and in figure 2d with $\beta=2.0$. In all cases, the queen happens to be the top ranked individual in each of the $100$ configurations. As from our data set it is evident that the interaction probabilities are heterogeneous in nature, we expect a non-zero value of $\beta$ for our species. As $\beta$ increases gradually from $0$, the individuals with higher winning abilities tend to interact more often with the others and the percentage of dominance in the first columns gradually increase from $14\%$ for $\beta=0$ to reach $36\%$ for $\beta=2.0$. Since for both the \textit{R. marginata} and \textit{R.cyathiformis}, the first column of dominance percentage is around $30\%$, we expect our desired $\beta$ would be around $2$, so we keep $\beta=2.0$ for the rest of the variation.
 
Next we examine the behaviour of the model for various values of the monotonicity parameter $\alpha$, keeping $\beta=2.0$ and $w=0$. For three different values $\alpha=0$, $\alpha=0.5$ and $\alpha=1.0$, the results we get are depicted in the figure 2d, 2e and 2f. In figure 2d, for $\alpha=0.0$, following the line of figure 2c, most of the dominance behaviour is shown by the top ranked individual who is also the queen in all of the $100$ configurations. Here the interaction probabilities of the individuals are monotonically connected with their strengths; higher strengths lead to higher interactions. As we increase the value of $\alpha$ from $0$, this monotonic relationship breaks and the probability of the existence of individuals with low strength but higher interaction increases. For $\alpha=0.5$ (figure 2e), we get some low ranked individuals who show more subordinate behaviour than others, a characteristic that we have observed in our study species. Therefore we expect for our species, the value of $\alpha$ would be around $0.5$. But in this case also, in $97$ out of $100$ configurations, the queen retains the top position in the hierarchy. For $\alpha=1.0$ (figure 2f), we get an inverse monotonous relationship between strength and interaction probabilities, where higher strengths lead to lower interaction probabilities. We get almost a mirror image of the pattern in figure 2d; the lowest individual in the hierarchy shows most of the subordinate behaviour, dominance behaviour is shared among all the individuals almost equally except for the lowest individual, and only in $1$ out of $100$ cases, the queen retains the top position. Since the $\alpha$ values are subtracted from the strength function $x_i$ whose range is between $0$ and $1$, we vary $\alpha$ also from $0$ to $1$. In both the ends of the scale, i.e., for $\alpha=0.0$ and $\alpha=1.0$, we get a monotonous relationship between $x_i$ and $p_i$; in between these extremities, the monotonicity breaks.

The effect of non-interacting individuals on the model is shown in the next three figures. Here we keep $\alpha=0$ and $\beta=2$ fixed for all three cases and vary $w$. As we keep on increasing the percentage of non-interacting individuals to $10\%$, $30\%$ and $50\%$, the total dominance-subordinate interactions are shared by the remaining individuals, so the percentage of interactions for at least one of them also increases. We can see the effect clearly in the first dominance bin (figure 2g, 2h and 2i). The percentage of cases where the queen holds the top position also decreases gradually as $90$, $67$ and $49$ respectively, since the chance that the queen becomes a non-interacting individual also increases gradually. Table 2. summarizes different combinations of parameter values used in the shown figures.
\begin{figure}
\begin{center}
\begin{tabular}{cc}
\includegraphics[width=8cm]{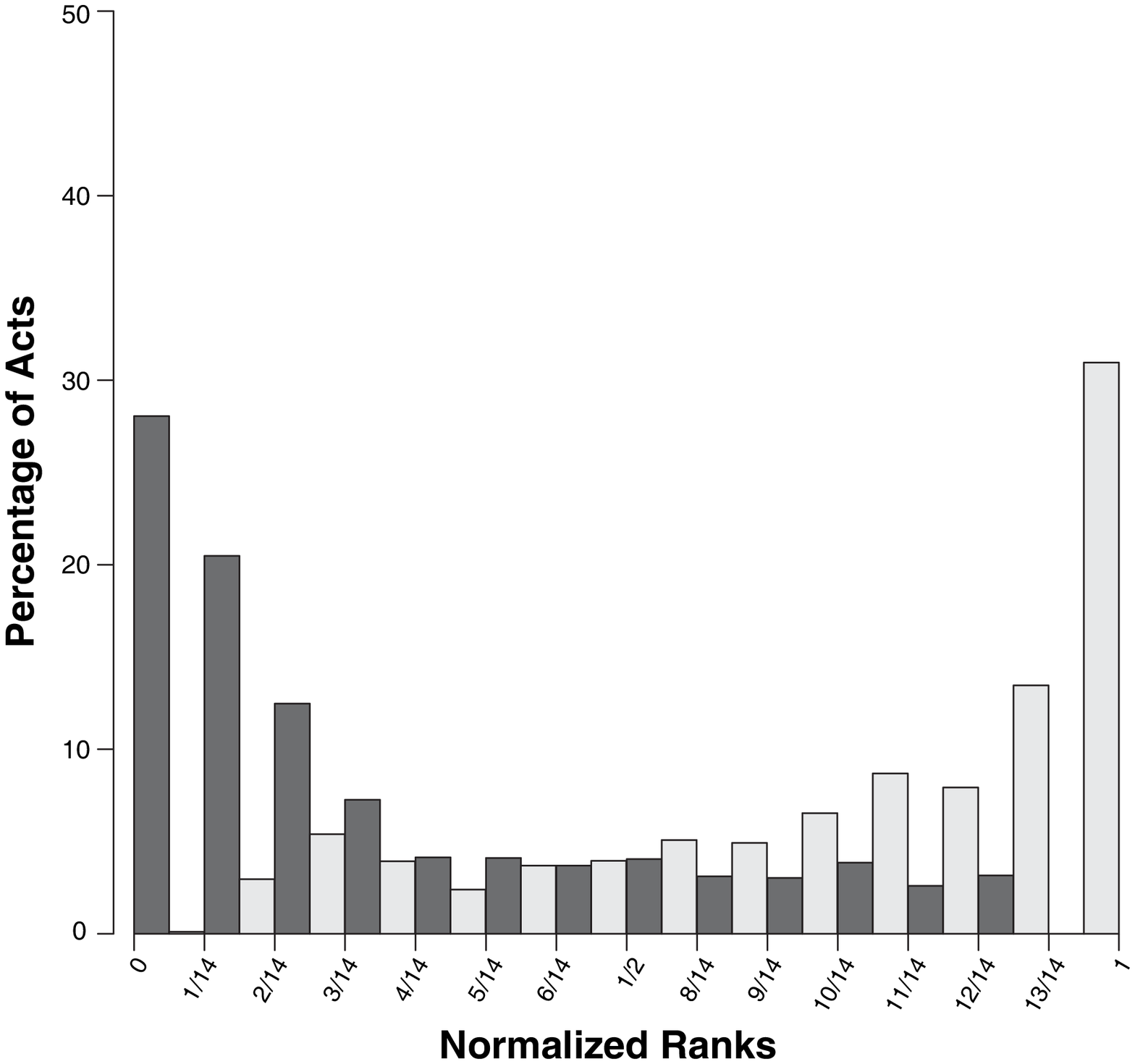}\\
\includegraphics[width=8cm]{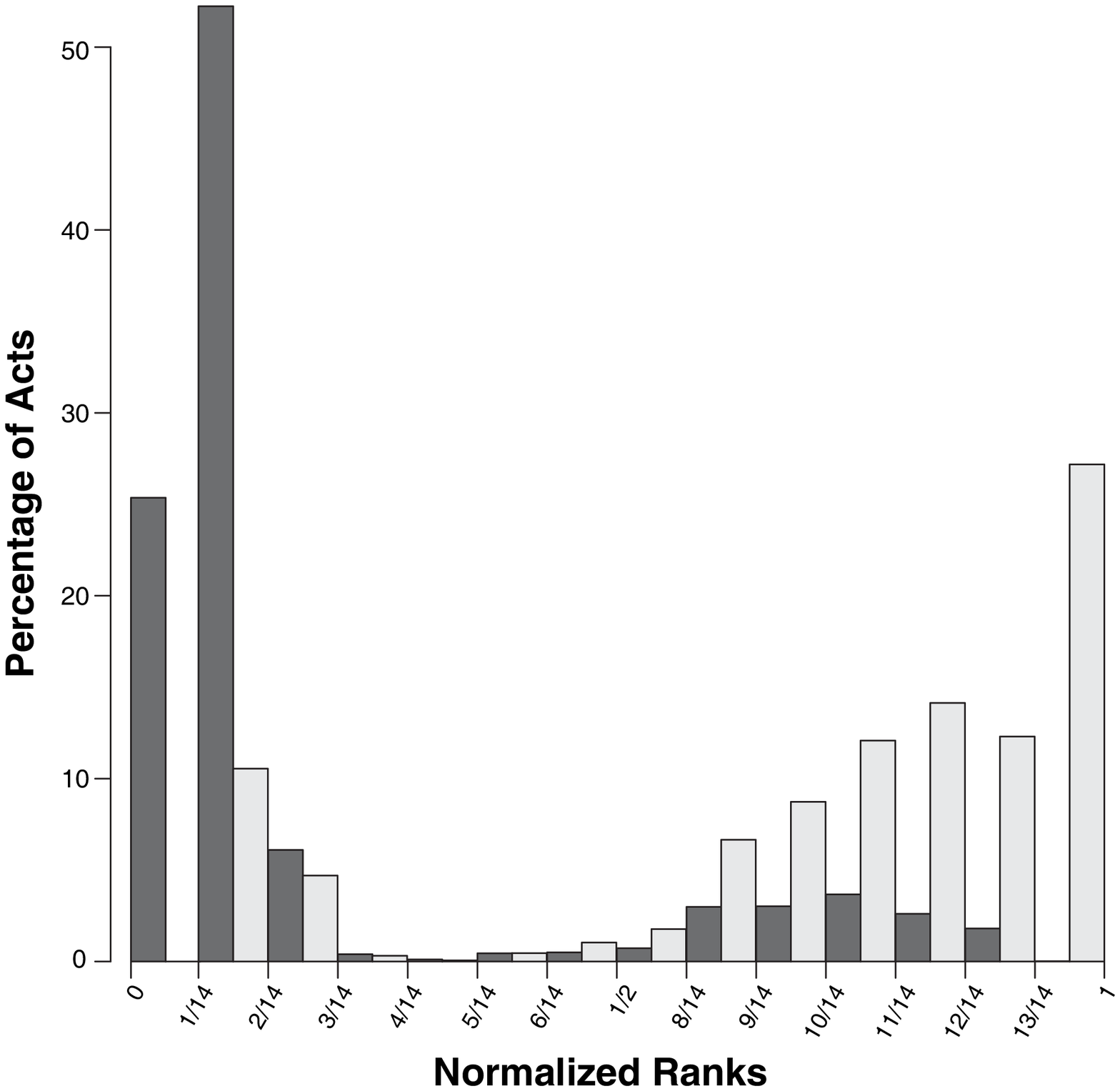}
\end{tabular}
\end{center}
\caption{Percentage of dominance acts (black bars) and subordinate acts (gray bars) plotted against the normalized ranks obtained by averaging over 100 configurations of 100 interactions with parameters a) $\beta=2$, $w=4$, $\alpha=0.43$ and $p_Q=av(p_i)/N$ and b) $\beta=2$, $w=50$, $\alpha=0.31$ and $p_Q=av(p_i)$.}
\end{figure}

\subsection{The Two Different Strategies}
What will happen if the queen changes her strategy slightly? To investigate this situation, we change by hand the $p_i$ of the individual with highest $x_i$ , i.e., of the queen, to a small value, say $p_Q=av(p_i)/N$ where $av(p_i)=(\Sigma p_i)/N$. For $\beta=2$, $\alpha=0.50$ and $w=0\%$, we obtain a pattern qualitatively very similar to figure 1a where like \textit{R. marginata}, in most of the cases the queen does not hold the top position in the hierarchy. The value we set for $p_Q$ is arbitrarily chosen to be very close to $0$ ($\sim$ or $< 0.01$), so that the queen becomes almost non-interacting in terms of dominance behaviour, a feature which is common in \textit{R. marginata}. As we take the value of $p_Q$ away from $0$, the queen’s interactions increase and her chance of holding the top position in the hierarchy also increases in turn. If we keep the value of $p_Q$ at such a small but non-zero value, that the queen gets involved in a moderate number of interactions, for example $p_Q=av(p_i)$, ($>0.01$ and $<0.1$, all other parameters being the same), we get a pattern qualitatively similar to figure 1b. Here, the queen does not always show maximum dominance interactions but mostly holds the top position in the hierarchy, the feature that is common in \textit{R. cyathiformis}. The differences in patterns introduced by the change in queen’s strategy are in general similar in nature for all parameter values.

We search through the parameter space for quantitative similarity with statistical significance. For $\beta=2$, $\alpha=0.43$, $w=4\%$ and $p_Q=av(p_i)/N$, we obtain the pattern seen in figure 3a. We get a different pattern with $p_Q=av(p_i)$, $\beta=2$, $\alpha=0.31$ and $w=50\%$ (figure 3b). The differences between figure 3a and figure 3b for both the dominance and subordinate patterns are found to be significant at $95\%$ confidence level by Kolomogorov-Smirnov two-sample test. The effect sizes (Cliff’s delta) of $0.44$ and $0.05$ were measured for dominance and subordinate behaviours respectively and all the Cohen’s $d$ indices are furnished in Table 1. In the first case we find that in $23$ out of $100$ cases the queen holds the top rank in the hierarchy (figure 3a). We use the Kolomogrov-Smirnov test for goodness of fit \cite{Sok.Rof.94} and find that the differences between figure 3a and figure 1a for both the dominance and subordinate patterns are non-significant at $95\%$ confidence level (Cliff’s delta for dominance and subordinate behaviours are $0.04$ and $–0.03$ respectively, Cohen’s indices are shown in Table 1). It is also worth noting that in the \textit{R. marginata} data set $14\%$ of the total individuals were non-interacting and from a different analysis of $100$ colonies of \textit{R. marginata}, we know that in $16$ colonies the queen was the top ranked individual \cite{Bh.inprep}. In the second case we find that in $99$ out of $100$ cases the queen holds the top position in the hierarchy (figure 3b). We again use the Kolomogrov-Smirnov test for goodness of fit and find that the differences of the patterns in figures 3b and figure 1b are both non-significant at $95\%$ confidence level (Cliff’s delta for dominance and subordinate behaviours are $-0.10$ and $–0.04$ respectively, Cohen’s indices are shown in Table 1). In the \textit{R. cyathiformis} data set, there were $40\%$ non-interacting individuals and in $100\%$ colonies, i.e., $9$ out of $9$, queens were at the top position of the hierarchy.

Our thorough investigation of the parameter space reveals that, for a range of combinations of the three parameters $\alpha$, $\beta$ and $w$, we get the non-significant difference between our model results and the real data. We show the regions where we get $p>0.05$ in Kolomogorov-Smirnov test for goodness of fit in figure 4a for \textit{R.marginata} and in figure 4b for \textit{R.cyathiformis}. We found that, near the observed values of $w$ ($14\%$ for \textit{R.marginata} and $40\%$ for \textit{R.cyathisformis}) and the approximately estimated values of $\beta$ (explained in the next section), there is a range of non-zero $\alpha$ values for which the model holds. We investigate the sensitivity of the parameters of the model using effect sizes also. We show the regions where we get a small mismatch between the model results and the observed data ($|$Cliff’s delta$|<0.1$) in figure 4c for \textit{R.marginata} and in figure 4d for \textit{R.cyathiformis}. We observe that, by using the effect size, we get bigger regions of validity those are supersets of the regions indicated by Kolomogorov-Smirov test. 
\begin{figure*}
\begin{center}
\includegraphics[width=6cm]{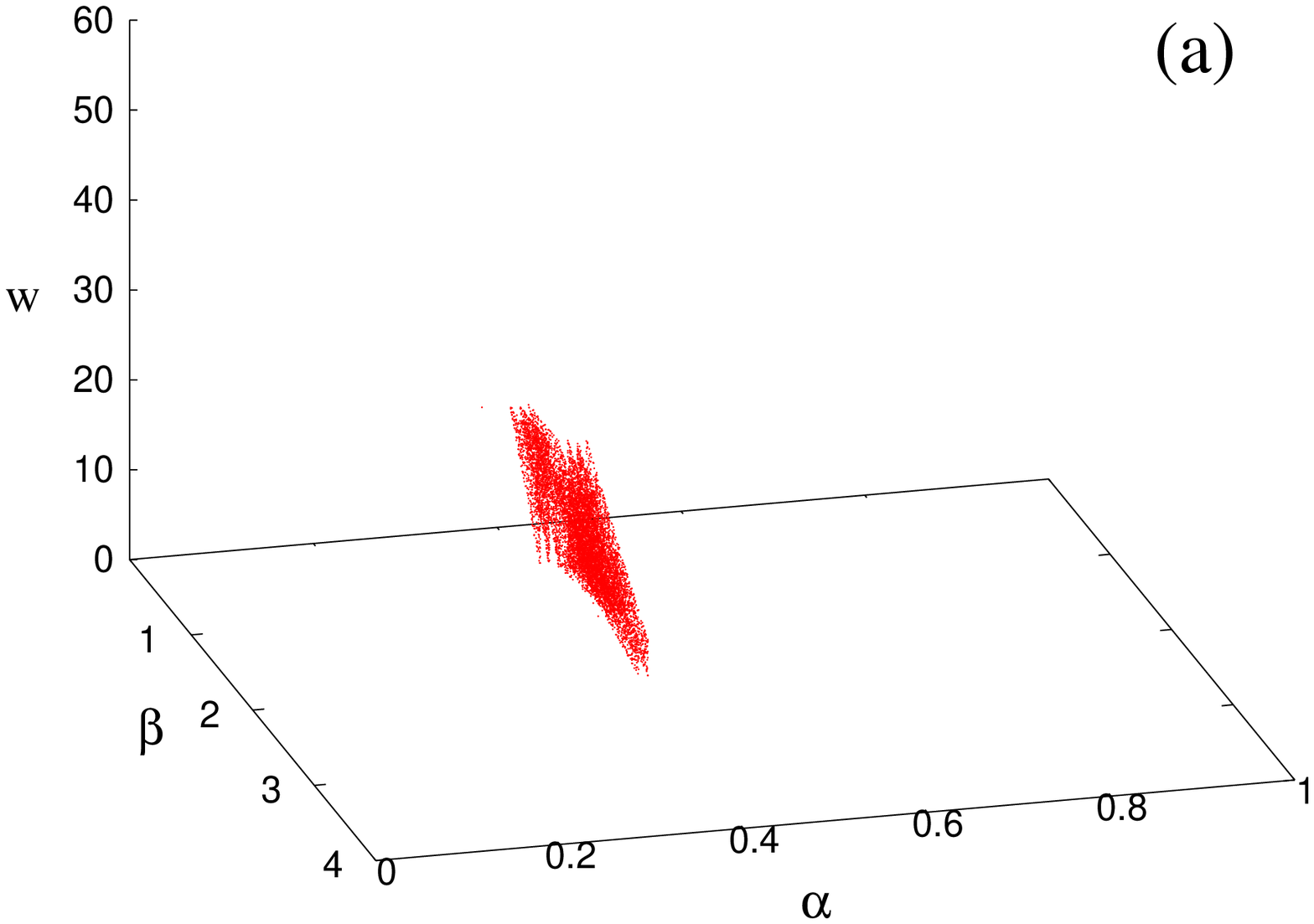}
\includegraphics[width=6cm]{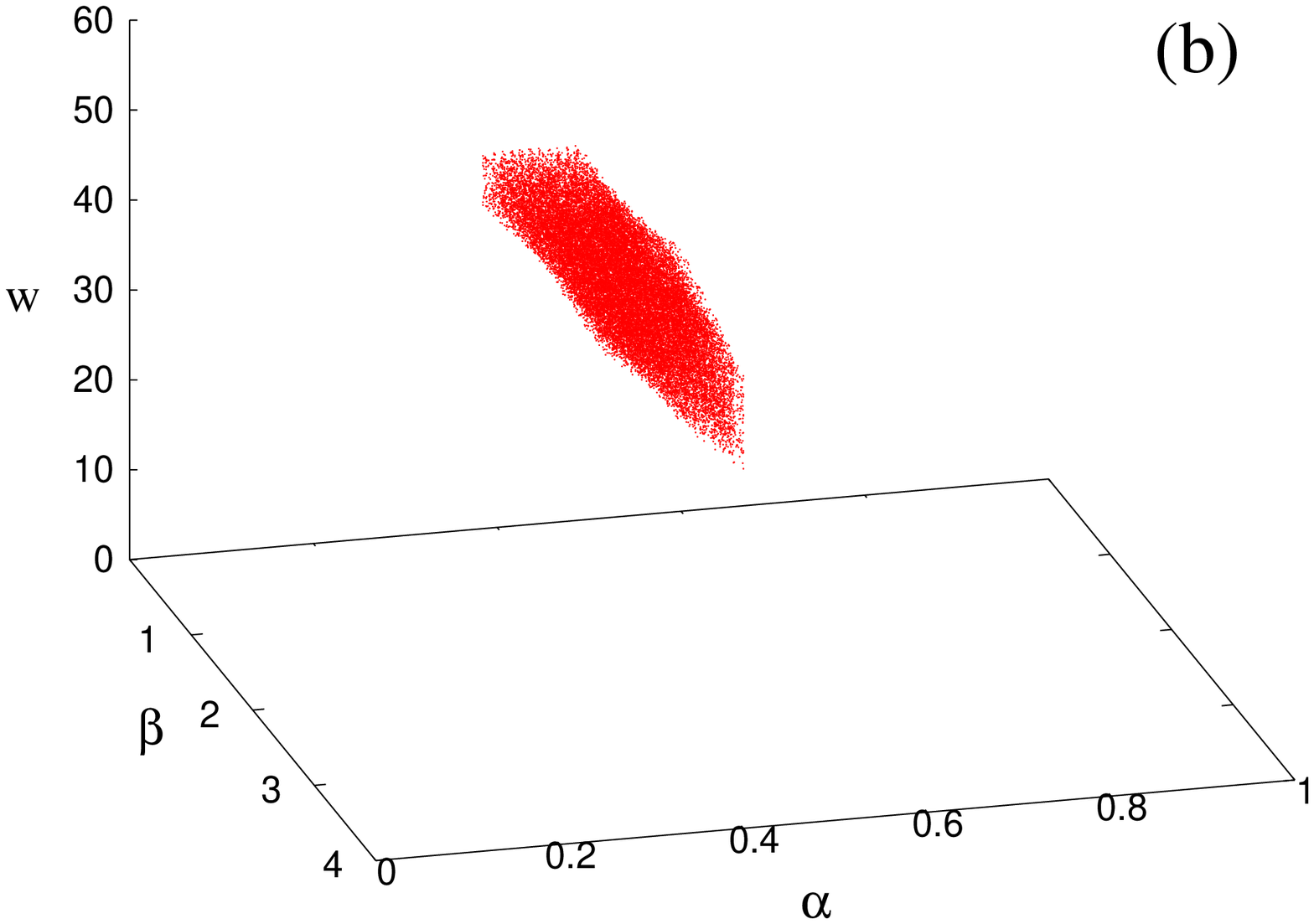}\\
\includegraphics[width=6cm]{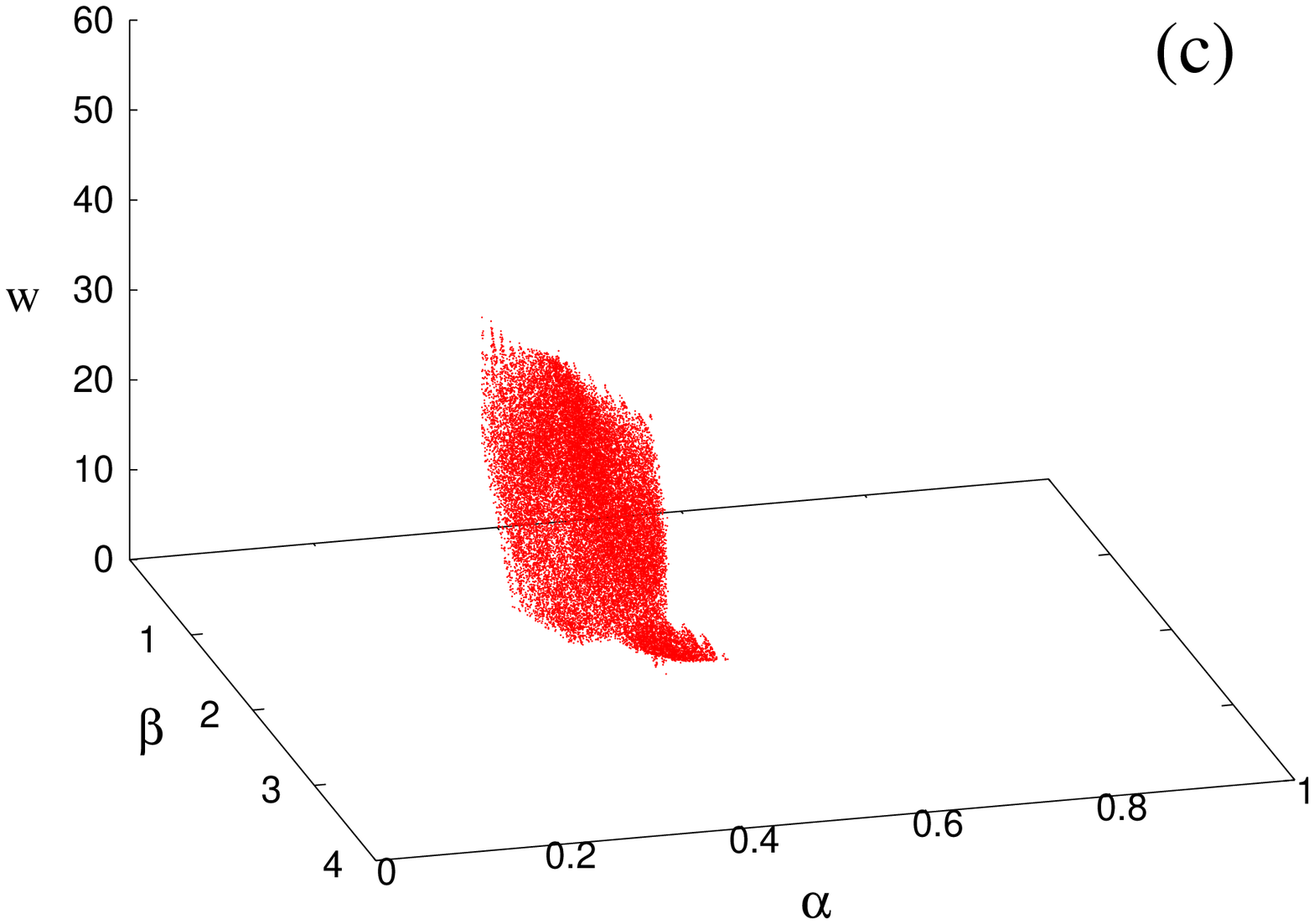}
\includegraphics[width=6cm]{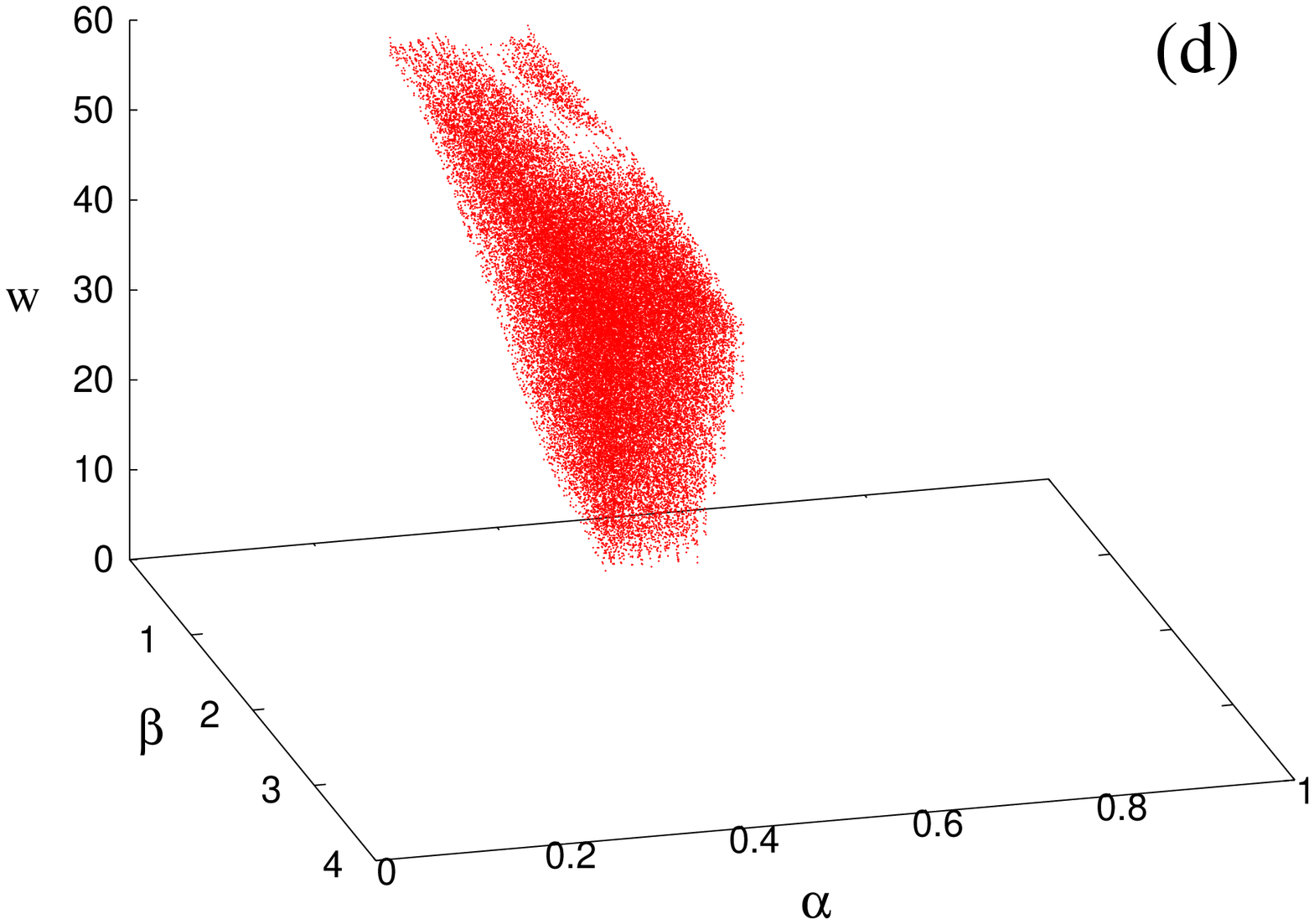}
\end{center}
\caption{The sensitivity analysis for the parameters of the proposed model. The regions indicate the combination of parameters for which the model results are indistinguishable from the observed data. The criteria for testing the validity: $p>0.05$ in Kolomogorov-Smirnov test for goodness of fit, a) \textit{R.marginata}, b) \textit{R.cyathiformis}. Also: $|$Cliff’s delta$|<0.1$ as a measure of effect size, c) \textit{R.marginata}, d) \textit{R.cyathiformis}.}
\end{figure*}

\subsection{The Distributions of Interaction-probabilities}
What could be the significance of the functional relationship of the strength $x_i$ and interaction probabilities $p_i$? In other words, having the above-mentioned relationship between $x_i$ and $p_i$, what could we predict about the interaction probabilities $p_i$? Since $x_i$'s are taken from a uniform random distribution, the probability density function $P(x)$=constant. And since we have considered $p_i$ as a function of $x_i$ and for $\beta=2$, $p_i \sim (x_i-\alpha)^2$ , one could write $x_i \sim p_i^{1/2}$ , therefore:\\ 
$P(p)dp = P(x)dx$\\ 
or, $P(p) = (dx/dp)P(x)$\\
or, $P(p) \sim (dx/dp) = (1/2)p^{(1/2-1)}$\\
or, $P(p) \sim p^{-0.5}$\\   
So, one could expect that the density function for the interaction probability $P(p)$ should fall as a power-law with an exponent of $-0.5$. We constructed the distributions of interaction probabilities for both the species from the data (figure 5). We tried to fit a non-linear regression function $y=Ax^{-B}$ to the plots. In figure 5a, for \textit{R. marginata}, we get $A=0.94$ and $B=0.69$, both $A$ and $B$ are significant at $95\%$ confidence level (analysis of variance test using $F$-statistics \cite{Zar.09}). In figure 5b, for \textit{R. cyathiformis}, we get $A=0.84$ and $B=0.70$, where $B$ is significant at $95\%$ level, but $A$ is not. These results qualitatively give justification for our assumptions about the functional relationship between the strength function and the interaction probability. The power-law distribution in interaction probability suggests that one can get smaller values (very close to zero) with a greater probability and higher values (in this case, close to $0.5$) with a small but finite probability.  

\subsection{\textit{Polistes dominulus}}
The social organization of the temperate paper wasp \textit{P. dominulus} has been studied in detail over many years. It is considered as a typical primitively eusocial wasp species, lacking morphological distinction between the queen and worker castes. The queen in \textit{P. dominulus} holds the topmost rank in the dominance hierarchy, as in \textit{R. cyathiformis}. But unlike in \textit{R. cyathiformis}, the \textit{P. dominulus} queen always shows most of the dominance behaviour and all the other individuals share the subordinate behaviours almost equally \cite{Pardi.48, Ther.etal.89, Therau.etal.92}. High ranked individuals other than the queen usually indulge in nest building and brood care while the others take up the job of foraging \cite{Therau.etal.92, Therau.etal.90}. The dominance hierarchy is therefore coupled with the organization of labour in the colony, which is generally regarded as an important factor in the evolution of eusociality \cite{Ost.Wil.78}. We were interested in checking if slight changes in individual strategies in our model could also give rise to the patterns similar to those seen in the social interactions of \textit{P. dominulus}, which are different from those observed in \textit{R. marginata} and \textit{R. cyathiformis}. 

We have seen in figure 2d that, with $\alpha=0.0$, $\beta=2.0$ and $w=0$, the queen shows the most dominance behaviour ($36.1\%$) and also holds the top position in the hierarchy in each of the $100$ configurations. Subordinate behaviours are shared almost equally by all the individuals except the queen. This is the qualitative pattern of dominance in \textit{P. dominulus} colonies reported by Pardi \cite{Pardi.48} and Theraulaz et al. \cite{Ther.etal.95}. However, it should be noted that a different dominance index was used by these authors, and all the data were from observations of pre-emergence nests while ours are of post-emergence nests. We believe that our model would also be applicable for post-emergence nests of \textit{P. dominulus}.

\section{Discussion}
\subsection{Strategies in \textit{R. cyathiformis}}
Our analysis suggests that from a typical primitively eusocial species, where the queen holds the top position in the dominance hierarchy and also shows the most dominance behaviour, more complex societies could evolve by changing the queen's as well as the workers' strategies. The queen could slow down her interaction rate to a moderate value but still remain at the top position of the hierarchy if she directs some of her aggression towards a single individual, the second in rank, who in turn would dominate the others in the colony. In case of loss of the queen, the individual who was second in rank and who is also the second strongest individual (in terms of $x_i$), could become the new queen and later eventually slow down her aggression. This strategy is likely to be observed in \textit{R. cyathiformis}, where the queen holds the topmost rank in the dominance hierarchy of the colony but the second ranked individual often shows a considerable amount of dominance, sometimes more than the queen. When the queen dies or is experimentally removed, the second ranking individual steps up her aggression and becomes the potential queen of the colony, which is reminiscent of similar situations in \textit{R. marginata} \cite{Kar.Gad.03}. Hence it is likely that the potential queen in \textit{R. cyathiformis} eventually becomes the new queen of the colony, a situation that we have not yet demonstrated experimentally.
\begin{figure}
\begin{center}
\includegraphics[width=8cm]{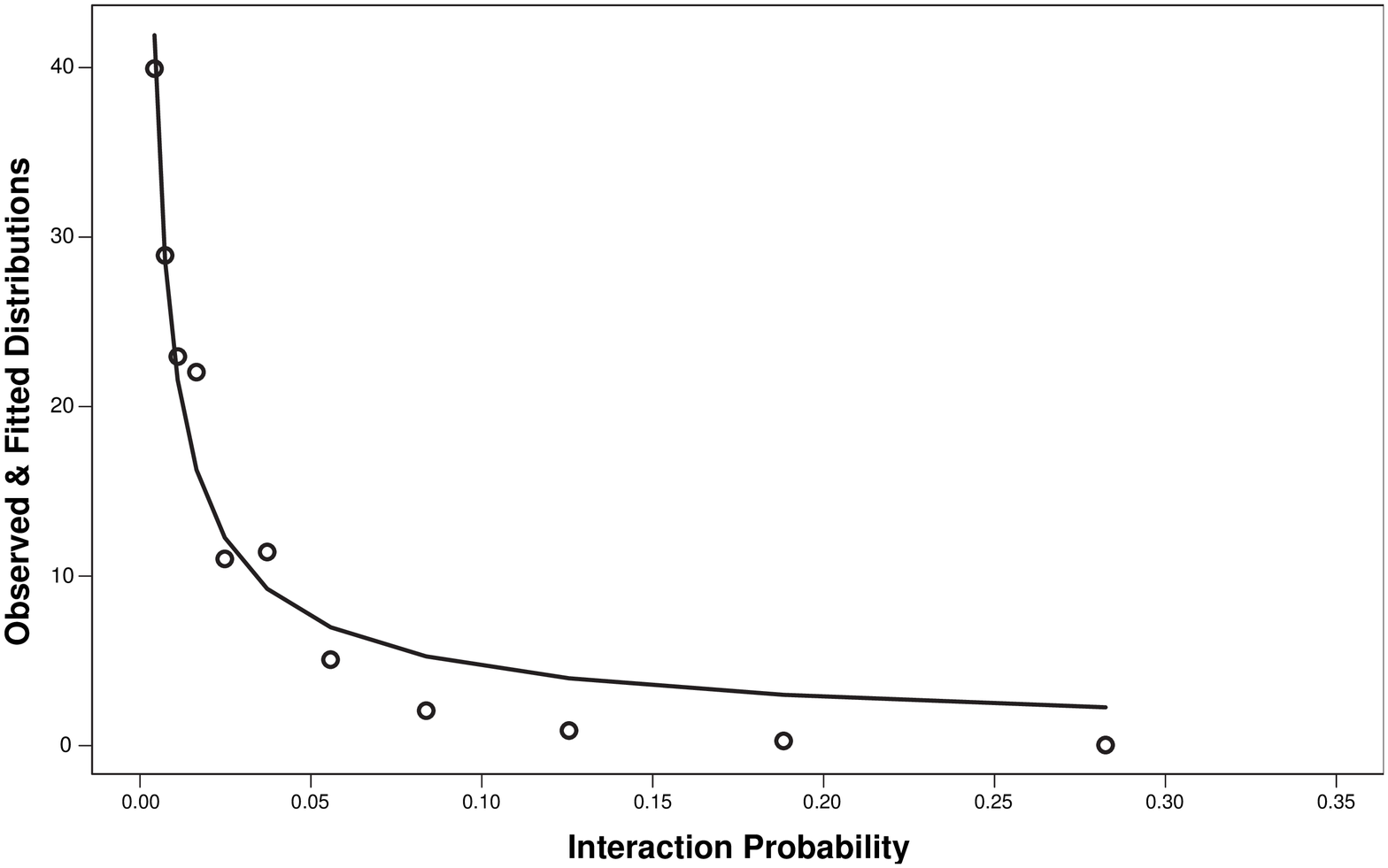}
\includegraphics[width=8cm]{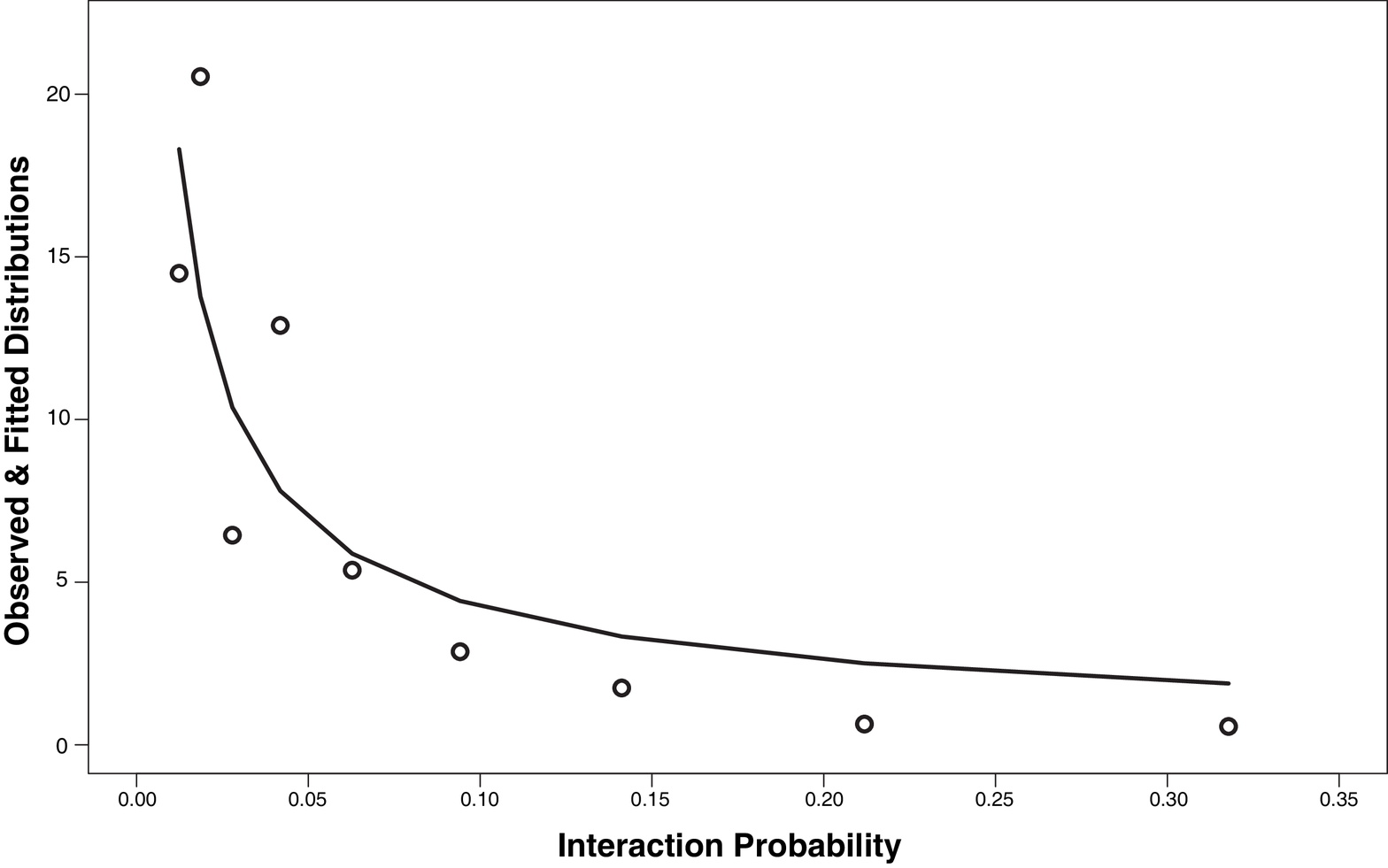}
\end{center}
\caption{Probability density distribution (open circles) for the interaction probabilities for 9 colonies of a) \textit{R. marginata} and b) \textit{R. cyathiformis}. The solid curve signifies the fitted distributions of the form $y=Ax^{-B}$.}
\end{figure}

\subsection{Strategies in \textit{R. marginata}}
In \textit{R. marginata} colonies, no attempts at egg-laying by the workers has ever been recorded and all workers have much poorly developed ovaries as compared to the queen \cite{Chandra.Gadagkar.91, Gadagkar.01}. However, in \textit{R. cyathiformis} colonies, the queen is not always the sole egg layer, occasional egg-laying by one or a few workers has been recorded \cite{Gadagkar.01}. Thus the reproductive threat to the queen is likely to be low in \textit{R. marginata} as compared to \textit{R. cyathiformis}, a condition which has definitely been achieved through a more advanced and efficient control system. The queen in \textit{R. marginata} is actually known to use pheromones to regulate worker reproduction \cite{Sum.etal.08, Bh.etal.10}, while such a pheromone is not yet known in \textit{R. cyathiformis}. It is possible that the \textit{R. marginata} queen has adopted the strategy of slowing down her dominance interaction rate to an even lower value as compared to \textit{R. cyathiformis}.  Since the \textit{R. marginata} queen does not require to expend energy in dominance interactions, she can use more energy for reproduction as well as for production of pheromones. However, though the \textit{R. marginata} queen does not occupy the topmost position in the dominance hierarchy, she does not lose the ability to be aggressive, and continues to be the strongest individual (in terms of $x_i$). She can resort back to aggression if required for maintaining her status in the colony \cite{Saha.etal.12}. If the \textit{R. marginata} queen is lost or removed from the colony, the second strongest individual, who probably was also using the less-interaction strategy, takes up the queen’s job. We observe one of the workers to become extremely aggressive on death or removal of the queen, but within a few days she develops her ovaries and begins egg-laying \cite{Prem.etal.96, Sumana.Gadagkar.03}. During this period she also gradually reduces her aggression and eventually becomes a meek and docile queen, channeling her energy towards reproduction. 

\subsection{Non-interacting Workers}
Considering different percentages of individuals who do not take part in any dominance-subordinate interactions, we obtained two distinct patterns (figure 3), with $4\%$ and $50\%$ non-interacting individuals respectively. In the \textit{R. marginata} colonies, we find that $14\%$ of the individuals did not interact, while in \textit{R. cyathiformis} this value was $40\%$ (difference is significant at $95\%$ confidence level, 2-proportion $Z$ test). We speculate that the \textit{R. cyathiformis} queen, being the strongest individual in the colony, uses some of her aggression towards the second strongest individual, who might use her aggression to directly recruit workers for foraging and other colony maintenance activities. Hence the interactions are limited among the few individuals who are directly involved in work regulation. In \textit{R. marginata}, dominance behaviour is used by the workers to regulate each others' foraging activities; the frequency of dominance behaviour in the colony decreases with decreased hunger levels, and increases with increased hunger levels in the colony \cite{Bruyndonckx.etal.06, Lamba.etal.07}. Workers in \textit{R. marginata} can be differentiated into three behavioural castes - sitters, fighters and foragers. While the foragers are involved in foraging activities, the fighters are the ones showing maximum aggression, and the sitters typically belong to the non-interacting group \cite{Gadagkar.01, Gadagkar.Joshi.83}. Since work regulation is achieved through a decentralized system of recruitment, colony maintenance activities are performed through a two-tiered system in which some individuals use aggression to make others work, so that interactions are not limited to only a few individuals, as in \textit{R. cyathiformis} \cite{Kar.Gad.02}. Hence the low percentage of non-interacting individuals is explained by the change in the strategy of work regulation.
\begin{figure*}
\begin{center}
\includegraphics[width=12cm]{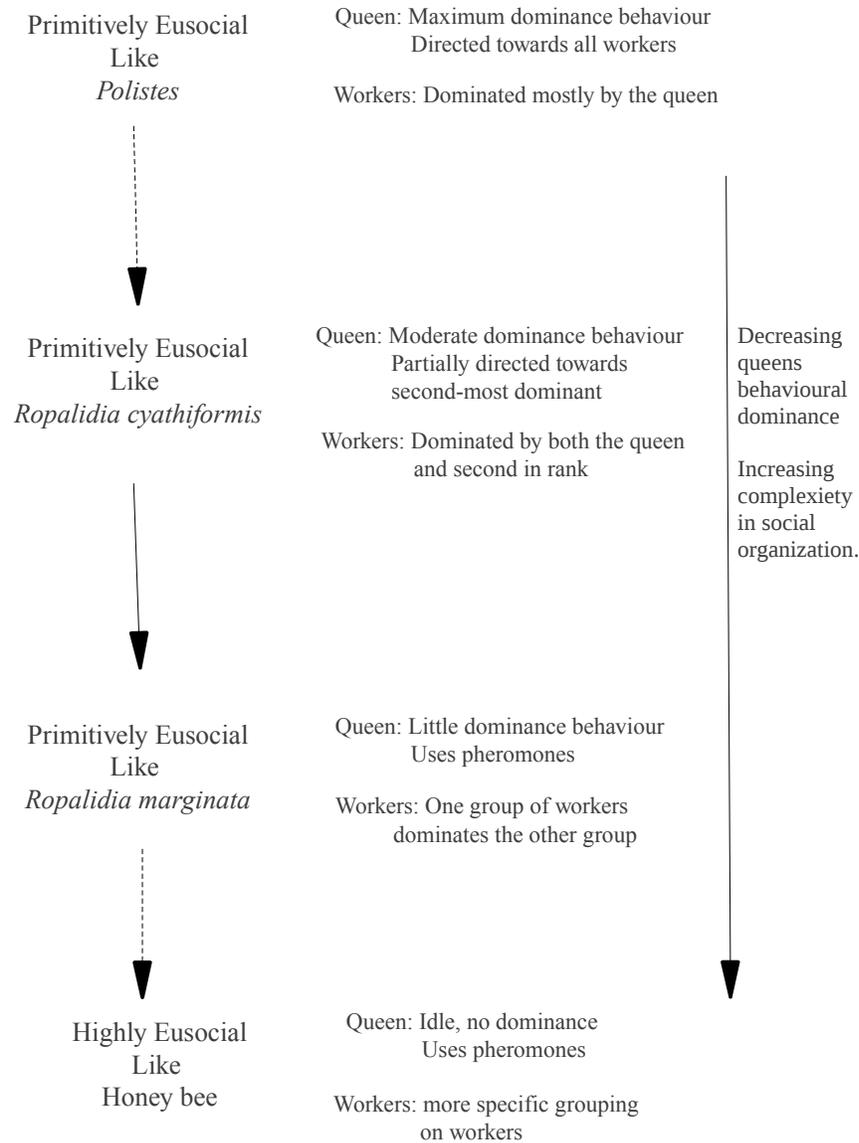}
\end{center}
\caption{A schematic diagram tracing the plausible evolutionary pathway.}
\end{figure*}

\subsection{Conclusions}
Through our model, we have proposed a common mechanism, with simple changes which could give rise to the observed dominance-subordinate patterns in both the primitively eusocial species \textit{R. cyathiformis} and \textit{R. marginata}. At one end of the model we have \textit{Polistes}-like patterns, where the queen holds the top position in the hierarchy and also shows most of the dominance interactions. A simple reduction in the queen's interaction allows the evolution of a hierarchical control system where the queen still holds the top position in the hierarchy but does not necessarily show most of the aggression in the colony. She directs some of her aggression towards the second ranking individual, who in turn helps her to control worker activities in the colony. These changes could lead to the evolution of a social system like \textit{R. cyathiformis}. A further reduction of the queen's interaction turns the control system towards a more decentralized one, where worker activities are controlled by the workers themselves; while the queen, who also evolves a pheromone to signal her presence to her workers, does not hold the top position in the hierarchy any longer. These changes lead to the evolution of a social organization like that of \textit{R. marginata}. The changes required to go from an \textit{R.cyathiformis}-like system to an \textit{R.marginata}-like system may be harder to achieve than the changes required to go from a \textit{Polistes}-like system to an \textit{R.cyathiformis}-like system, given that the former necessitates the physiological changes required for the production and perception of pheromones. But once achieved, it would allow the colony to increase its size by producing more workers and also allow the queen to channel most of her physical energy to reproduction, eventually becoming a morphologically large egg-laying machine, as in the highly eusocial species like ants and the honeybees. Using our model, we have traced out a plausible evolutionary pathway through which more complexities in social organization could have evolved (figure 6).

\subsection*{Future Directions}
Although our model is fundamentally similar to the correlational models discussed in section 3, there are some important differences. We have used a mathematical relationship by which the strengths of the individuals are connected to the probabilities through which they interact. The predictions about the interaction probabilities are consistent with the experimental observations for the concerned two species. It would be interesting to know if the knowledge of interaction probabilities could lead to the formation of the expected dominance-subordinate pattern, i.e., if a correct prediction is possible for the dominance-subordinate pattern from the distribution of interaction probabilities. Existing dominance data from other group-living species could be re-examined along this line. A successful prediction would certainly substantiate the model to a large degree. Another important aspect is the strategy of the individual with highest strength, a change in which could vary the dominance profile a lot. It would be very interesting to examine the effect of such changes in animal societies of higher complexities where other complicated factors could be present. We are also encouraged by the successful predictions made by the self-organized models in case of \textit{P. dominulus} \cite{Bon.etal.99}. Our lab has recently carried out some experiments to validate the assumptions of self-organized models and we are now trying to verify the results in terms of reinforcement of dominance abilities. 

\section{Acknowledgements}
This study was supported by the Department of Biotechnology, Deaprtment of Science and Technology, Ministry of Environment and Forests and the Council for Scientific and Industrial Research, Government of India. The data analysis, modeling, simulations and statistical tests were performed by AKN. The results were interpreted by AKN and AB. Behavioural observations on Rm were carried out by AS and on Rc by SAD. The paper was co-written by AKN, AB and RG, and RG supervised the whole work. AKN wishes to thank Dr. Kunal Bhattarcharya, BITS, Pilani, India for his valuable feedbacks. All experiments reported here comply with the current laws of the country in which they were performed. The analysis are performed by using the statistical environment `R' \cite{R.08}.

\end{document}